\documentclass[sigconf,screen,nonacm]{acmart}
\AtBeginDocument{%
  \providecommand\BibTeX{{%
    \normalfont B\kern-0.5em{\scshape i\kern-0.25em b}\kern-0.8em\TeX}}}

\usepackage{calc}
\usepackage{subfigure}
\usepackage{multirow}

\definecolor{NoteBlue}{RGB}{0,0,191}

\definecolor{TodoRed}{RGB}{225,63,63}

\definecolor{ControversialGreen}{RGB}{0,127,0}

\definecolor{UnfinishedBlue}{RGB}{0,127,127}

\definecolor{KKOrange}{RGB}{255,103,2}

\newcommand{\fref}[1]{Fig.~\ref{#1}}
\newcommand{\frefSub}[2]{Fig.~\ref{#1}{#2}}

\newcommand{\numerrate}[1]{\textbf{\texttt{(#1)}}}

\begin{document}

\title{Conducting User Studies in Augmented Reality - Challenges and Obstacles}
\title{Experiences with User Studies in Augmented Reality}

\author{Marc Satkowski}
	\orcid{0000-0002-1952-8302}
  \affiliation{%
    \department{Interactive Media Lab Dresden}
    \institution{Technische Universität Dresden}
    \city{Dresden}
    \country{Germany}
  }
	\email{msatkowski@acm.org}
\author{Wolfgang Büschel}
	\orcid{0000-0002-3548-723X}
    \affiliation{%
        \department{Interactive Media Lab Dresden}
        \institution{Technische Universität Dresden}
        \city{Dresden}
        \country{Germany}
    }
	\email{bueschel@acm.org}
\author{Raimund Dachselt}
  \orcid{0000-0002-2176-876X}
  \affiliation{%
    \department{Interactive Media Lab Dresden}
    \institution{Technische Universität Dresden}
    \city{Dresden}
    \country{Germany}
  }
  \additionalaffiliation{%
    \department{Centre for Tactile Internet with Human-in-the-Loop (CeTI)},
    \institution{Technische Universität Dresden}
    \city{Dresden}
    \country{Germany}
  }
  \additionalaffiliation{%
    \department{Cluster of Excellence Physics of Life},
    \institution{Technische Universität Dresden}
    \city{Dresden}
    \country{Germany}
  }
  \email{dachselt@acm.org}

\renewcommand{\shortauthors}{Satkowski et al.}


% \input{content/abstract.tex}

  
\maketitle

% !TeX root = ../main.tex

% ##########################################################
% ######## Intro ###########################################
% ##########################################################
\section{Introduction}

% AR popular
The research field of augmented reality (AR) is of increasing popularity, as seen, among others, in several recently published surveys \cite{DeSouzaCardoso2020,Kim2018,Erickson2020,Fonnet2019,Merino2020,Dey2018}.
% Studies are important
To produce further advancements in AR, it is not only necessary to create new systems or applications, but also to evaluate them.
% UE is widely researched
One important aspect in regards to the evaluation is the general understanding of how users experience a given AR application, which can also be seen by the increased number of papers focusing on this topic \cite{Merino2020} that were published in the last years.
% UE essential in the market
With the steadily growing understanding and development of AR in general, it is only a matter of time until AR devices make the leap into the consumer market where such an in-depth user understanding is even more essential.
% Assessment and other factors
Thus, a better understanding of factors that could influence the design and results of user experience studies can help us to make them more robust and dependable in the future.
\par
\begin{figure*}
    \centering
    \includegraphics[width=\textwidth]{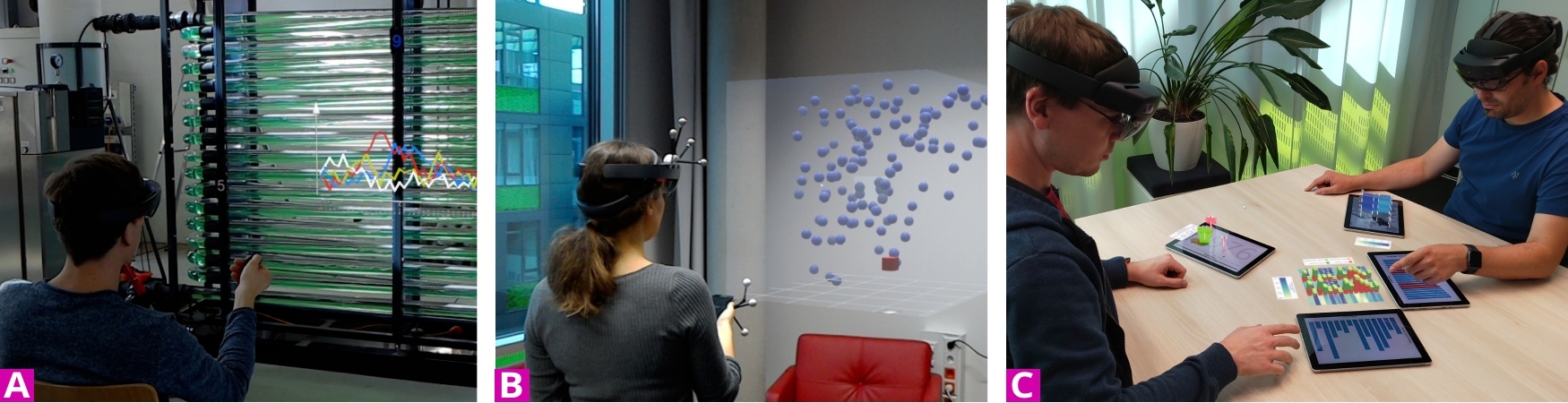}
    \caption{
        Different images from a few user study setups.
        % A - AR Background Perception
        \numerrate{A} \cite{Satkowski2021}
        A participants sits in front of a real-life algae reactor which is part of an industrial production plant.
        He has to analyze the shown visualization, while interacting via a Microsoft Clicker in his right hand.
        Tasks were answered orally.
        % B - Pan & Zoom
        \numerrate{B} \cite{Buschel2019a}
        In this study, we explored how mobile devices like a smartphone can be used to pan and zoom in 3D data spaces in AR.
        Here, the user holds the smartphone in her right hand and can use the spatial movement of the device to alter the view on the point cloud.
        Additionally, to unify their coordinate systems, the Microsoft HoloLens as well as the smartphone are equipped with tracking markers for a motion tracking system.
        % C - MARVIS
        \numerrate{C} \cite{Langner2021}
        Two persons, each wearing a HoloLens, analyze different data visualizations with a combination of four tablets.
        In this work, we presented a prototype that combines several mobile devices, HoloLenses, a motion tracking system, as well as an additional server application on a desktop PC.
    }
    \label{fig:study-images}
\end{figure*}

% Aim and content 
In this position paper, we describe three challenges which researchers face while designing and conducting AR users studies.
% Our experience
We encountered these challenges in our past and current research (see \fref{fig:study-images}), including papers that focus on
    perceptual studies of visualizations \cite{Satkowski2021,Buschel2019},
    interaction studies \cite{Buschel2019a}, and
    studies exploring the use of AR applications \cite{Buschel2021,Buschel2018} and their design spaces \cite{Langner2021}.
% !TeX root = ../main.tex

% ##########################################################
% ######## Study images ####################################
% ##########################################################

% ##########################################################
% ######## Challenges ######################################
% ##########################################################
\section{Challenges in AR User Studies}

User studies are a fundamental tool in HCI research.
However, the design process can be rather difficult and challenging, especially for AR systems.
Among those challenges, we want to name the following:
    \numerrate{1} more studies should be conducted in in-the-wild scenarios \cite{Dey2018,Fonnet2019,Satkowski2021,Merino2020},
    while a long-term user experience should be of concern \cite{Ens2021},
    \numerrate{2} generating insights can be difficult \cite{Merino2020}, 
    even with data collected under the best possible conditions \cite{Ens2021}, and 
    \numerrate{3} therefore a need for ``new evaluation methods that could capture more accurately the user experience in AR'' \cite{Kim2018}.
% Questions
In general, we will address two main questions that are connected to those challenges but also derived by our own experience:
\begin{itemize}
    \item [Q1] How can we ensure that the participants of a user study can easily give answers to various questions?
    \item [Q2] How can experimenters make sure that the participants solve the given tasks at hand in the correct manner and that the study prototype works as intended?
\end{itemize}
% Challenges
Based on those two questions, we will discuss the challenges of 
    the input capabilities of head-mounted display (HMD) AR devices,
    the communication between the participants and the experimenter, and
    the needed user study system setup in regards to the two previous challenges.
% HoloLens
Here we want to note, that we will only focus on optical see-through HMDs like the HoloLens v1 and the HoloLens 2, which we used in our own studies.
Therefore, some of the mentioned problems have to be assessed differently for other device types, like video-see through devices.

% ##########################################################
% ######## Input Capabilities ##############################
% ##########################################################
\subsection{Input Capabilities of AR}

% Intro - UX
Some parts of the user experience rely on the feeling how responsive and easy to use a system is.
However, the input capabilities of AR devices are rather challenging, both in regards to the general interaction, but also for completing specific study tasks and answering questionnaires.
% Questionnaires
The default way to fill out questionnaires, which we also often utilize in our AR studies, is to use pen and paper or a desktop computer.
% Questionnaire in AR
However, this procedure forces the participants to leave the designated study space and sometimes to remove the HMD as well.
To continue the study, it can be necessary to re-calibrate or restart the AR application which can introduce small alterations to the setup.
% AR capabilities
To minimize this problem, it would be possible to use the AR specific input modalities (in regards to the Microsoft HoloLens), like free-hand gestures, to fill out such questionnaires.
% List of Problems
Yet, this possibility has different problems associated with it:
\begin{itemize}
    % C1.1 - Inexperience
    \item [P1]
    Most of the invited participants in our user studies are inexperienced with AR and its input capabilities.
    Even an extended training session for free-hand gestures can only give them a shallow understanding, while increasing the overall study duration.
    % C1.2 - Recognition errors
    \item [P2]
    The general recognition of free-hand gestures is error prone, due to possible tracking errors.
    This can be amplified by certain subtleties of a gesture (e.g., hand has to be in the field of view of the HMD camera) that inexperienced participants are not aware of. 
    This can lead to not or wrongly recognized gestures.
    % C1.3 - Task completion time
    \item [P3]
    Greater fluctuations between answer quality and task completion time can appear for different tasks and participants while they use free-hand gestures as an input modality.
    Again, this can be linked to inexperience or possible recognition errors.
    % C1.4 - Not for all input types.
    \item [P4]
    Not all tasks and input types are suited for free-hand gestures, like text input.
    % C1.5 - Exhaustion of the participants
    \item [P5]
    An extended use of free-hand gestures can also lead to greater exhaustion, due to additional body, arm, or hand movement.
\end{itemize}
% Conclusion -> Solutions
In general, the usage of free-hand gestures allows the input needed for the answering of questionnaires, but seems rather slow, error prone, or too complicated for the participants.
% List of approaches
This is why different approaches can be used and explored to minimize the use of free-hand gestures for filling out questionnaires or the general interaction with an AR application:
\begin{itemize}
    % A1.1 - reducing set of needed interactions
    \item [A1]
    A reduction of the introduced interaction set can help to minimize the training needs and possible recognition errors.
    This is not always possible, especially for studies that explore complete AR systems and applications. 
    However, for studies that focus on perception or design choices, it can be a suitable solution.
    % A1.2 - other specialised input devices
    \item [A2]
    The usage of other input devices can leverage on other already known input modalities, like the Microsoft Clicker as a single mouse button (see \frefSub{fig:study-images}{a}).
    Again, this requires a quite simple interaction set, and in regards to the Clicker, can only be based on pointing.
    We used this, for example, in \cite{Satkowski2021}.
    % A1.3 - mobile devices
    \item [A3]
    Mobile devices, like smartphones or tablets, are another device type that can be used for interaction in combination with AR (see \frefSub{fig:study-images}{b}).
    In general, we are already quite familiar how to write text or fill out forms with such devices.
    Yet, the introduction of such devices brings other problems with it, which will be further described in \autoref{sec:c3}.
    % A1.4 - oral answering
    \item [A4]
    Participants could give feedback orally, like in an interview.
    We used this approach in \cite{Satkowski2021}.
    This can be easily realized, but makes it harder for the participants
        to think about the question, 
        to understand the question and its answer due to high noise levels in the environment, and
        to select an answer from a predefined set of answer (e.g., rating scales).
\end{itemize}

% ##########################################################
% ######## Participant-Experimenter Communication ##########
% ##########################################################
\subsection{Participant-Experimenter Communication}

% Intro - UX
Another essential part of user experience investigations is the ability 
    to talk with participants about the presented systems or designs,
    to verify and log what the participants do, and
    to communicate with the participants about certain parts of a system.
However, an AR HMD allows only one person to observe the virtual content, which generally prevents the experimenter from observing the study participants.
% Possible approaches
To address this, it is possible to extend the study application by different approaches:
\begin{itemize}
    % A2.1 - Additional logging client
    \item [A1]
    The AR application could be accompanied by an additional experimenter client on, e.g., a desktop computer, which logs specific activities and events made by the participant or the system. 
    We used this, for example, in \cite{Satkowski2021}.
    This allows the investigator to react to given predefined events but only gives a coarse understanding of the current user behaviour in the AR application and provides no visual feedback.
    %A2.2 - Streaming
    \item [A2]
    The experimenter could use the streaming capabilities of the AR HMD to see exactly what the participants currently do and see. 
    We used this in our study \cite{Langner2021}.
    However, this approach has several drawbacks, like
        a possible latency due to network issues,
        decreased performance on the AR device due to increased computational requirement (e.g., the HoloLens 2 will be limited to 30 fps with no antialiasing), and
        possible offsets in the position of virtual objects.
    %A2.3 - Additional HL
    \item [A3]
    The experimenter could also wear an AR HMD to see exactly what the participants see. 
    We used this, for example, in \cite{Buschel2021}.
    However, this approach requires at least two HMDs and could introduce other problems that we further describe in \autoref{sec:c3}.
\end{itemize}
In general, those additions allow for a better communication between the experimenter and the participant, but also have relative high costs in regards of development expenses and performance.
Lastly, with this increased need of preparation and implementation it takes far longer to create a study application.

% ##########################################################
% ######## Complexity of Multi Device Study Systems ########
% ##########################################################
\subsection{Complexity of AR Study Setups and Multi Device Applications}
\label{sec:c3}

% Intro - UX
AR applications are inherently more complex in comparison to desktop or smartphone applications, which makes it in general harder to analyze and study the overall user experience.
Such systems do not exist in isolation and are therefore connected with several other devices and objects all around them in the near environment.
% What can be connected to AR
Thus, it is only natural, that AR HMDs have to coexist with several other systems (see \frefSub{fig:study-images}{c}), like
    \numerrate{1} motion tracking systems for information about the position and rotation of objects,
    \numerrate{2} other mobile devices like smartphones and tablets with potential web applications,
    \numerrate{3} stationary devices like desktops or larger vertical displays,
    \numerrate{4} other AR devices that should be used in the same virtual space, or 
    \numerrate{5} other intelligent (e.g., washing machine, machines in industry 4.0) and normal everyday objects.
% Problems
However, this results in additional problems:
\begin{itemize}
    % C3.1 - Higher development costs
    \item [P1]
    With the increase of possible devices and objects that should be connected to an AR application, the costs for the development of such a system increases as well.
    This is mainly caused by the need for 
        additional testing,
        implementation of fallback mechanisms,
        synchronizing the local coordinate systems, and
        state synchronicity between each object.
    % C3.1 - In-the-wild
    \item [P2]
    Since AR applications aim to be adopted in real-life scenarios, such applications should also be able to be used and studied in those.
    However, 
        the availability of the needed infrastructure and
        interference by other present devices and networks
    could pose additional problems in such environments.
\end{itemize}
% Possible approach
One approach to reduce the complexity of such systems is to provide and use different frameworks that encapsulate specific functionalities.
Those can contain
    system behaviours (e.g., MRTK),
    visualizations (e.g., u2Vis \cite{Reipschlager:2020aa}),
    server communication, or
    interaction management.

% !TeX root = ../main.tex

% ##########################################################
% ######## Conclusion ######################################
% ##########################################################
\section{Conclusion}

% Recap
In this position paper we, presented three different challenges for AR user studies, specific problems linked to those, and possible approaches to counter them.
However, many of the mentioned problems are not easy to solve and the proposed solutions could introduce even more problems.
% UX
To be able to study how users perceive and experience such AR systems, it is important to be mindful about those challenges and to address them accordingly.
Such a robust system is especially important for user experience studies, since not only individual interactions, design concepts, or visualizations are investigated, but a whole application will be experienced by the users.
% Side effects
This is why possible side effects, like a decreased frame rate or faster exhaustion, can be quite critical and detrimental to the overall user experience.
% Future stuff
Since we only described a small subset of challenges connected to head-mounted AR devices, we want to note that there exist many more possible factors that can influence the users and their performance.
We hope that this position paper serves as a starting point for further discussions of the challenges in evaluating user experiences in Mixed Reality.

% !TeX root = ../main.tex

% ##########################################################
% ######## Acknowledgements ################################
% ##########################################################
\begin{acks}
    %General
    This work was funded in part by \grantsponsor{FD0001}{``Deutsche Forschungsgemeinschaft'' (DFG, German Research Foundation)}{https://www.dfg.de/en/index.jsp} 
    % RTG
    under grant number \grantnum{FD0001}{319919706/RTG2323} ``Conducive Design of Cyber-Physical Production Systems'',  
    % CPEC
    under project number \grantnum{FD0001}{389792660} as part of TRR~248 (see \url{https://perspicuous-computing.science}),
    % CETI
    under Germany's Excellence Strategy - EXC-2050/1 - Project ID \grantnum{FD0001}{390696704} - Cluster of Excellence ``Centre for Tactile Internet with Human-in-the-Loop'' (CeTI) of Technische Universität Dresden,
    % POL
    and under Germany's Excellence Strategy - EXC-2068 - \grantnum{FD0001}{ 390729961} - Cluster of Excellence ``Physics of Life'' of Technische Universität Dresden.
\end{acks}

\bibliographystyle{ACM-Reference-Format}
\bibliography{paper}

\end{document}